\documentclass{article}
\usepackage{latexsym}
\usepackage{color}
\usepackage{amsmath}
\usepackage{epsfig}
\usepackage{caption}
\usepackage{graphicx}
\newcommand{\ortala}[1]{\begin{center}#1\end{center}}

\linethickness{0.6pt}

\begin{document}

\ortala{\textbf{Skyrmion–Bimeron Transformation in Bilayer Chiral Magnets with Competing Magnetic Anisotropy}}
\ortala{\textbf{G\"ul\c{s}en Do\u{g}an, \"Umit Ak\i nc\i \footnote{umit.akinci@deu.edu.tr}}}
\ortala{\textit{The Graduate School of Natural and Applied Sciences, Dokuz
Eyl{\"u}l University, Tr-35160 {.I}zmir, Turkey}}
\ortala{\textit{Department of Physics, Dokuz Eyl\"ul University, TR-35160 Izmir, Turkey}}\textbf{}

\section{Abstract}

In this work, we investigate the emergence of topological spin textures in a ferromagnetically coupled bilayer chiral magnet by means of Monte Carlo simulations of a classical spin model including exchange interaction, Dzyaloshinskii–Moriya interaction, magnetic anisotropy, and an external magnetic field. To characterize the topology of the system, we construct a scalar chirality map in the $(K/J,h/J)$ parameter space.

Our results reveal several magnetic configurations, including labyrinth structures, skyrmion lattices, ferromagnetic states, and meron–antimeron crystal phases. In particular, we show that the transition from easy-axis to easy-plane anisotropy drives a continuous transformation from skyrmion textures to bimeron-type configurations. The bilayer geometry introduces an additional stabilization mechanism, where interlayer exchange coupling correlates the topological cores in the two layers and increases the energetic cost of defect collapse.

These findings provide a systematic topological texture map for bilayer chiral magnets and highlight coupled magnetic layers as a promising platform for stabilizing bimeron-type spin textures in nanoscale spintronic systems.

Keywords: \textbf{Skyrmion, Bilayer, Bimeron, Easy-Axis, Easy-Plane, Magnetic Anisotropy}

\section{Introduction}\label{introduction}

Skyrmions are topological entities that have long intrigued researchers in magnetism because they are associated with a range of physical phenomena. Magnetic skyrmions are compact, rotating topological defects within the magnetization configuration. A magnetic skyrmion is a local vortex of spin configurations in a magnetic material. The concept was first proposed in $1962$ by the nuclear physicist Tony Hilton Royle Skyrme, who coined the term skyrmion \cite{ref1}. Skyrmions were initially proposed in nuclear physics as specific solutions to a sigma model that represent baryons. Later, in $1989$, Bogdanov extended the idea to magnetic materials through theoretical predictions \cite{ref2}. Bogdanov and Hubert were among the first to theoretically investigate the role of the Dzyaloshinskii-Moriya interaction (DMI) \cite{ref3, ref4} in stabilizing skyrmion configurations in easy-axis magnetic systems and emphasized the importance of this term in such structures \cite{ref5, ref6}. Extending earlier theories, Bogdanov and R\"{o}\"{\ss}ler developed a phenomenological framework to describe chiral symmetry breaking in thin magnetic films and multilayers. Their work also predicted that skyrmions could arise from induced DMI in these systems \cite{ref7}. Then, R\"{o}\"{\ss}ler and collaborators extended these theoretical insights by showing that skyrmions could arise spontaneously as ground states in chiral magnetic metals, without the presence of external magnetic fields or structural defects \cite{ref8}. After being theoretically predicted for a long time, the existence of skyrmions was experimentally confirmed for the first time in $2009$ \cite{ref9}. Romming et al. successfully detected individual nanoscale skyrmions using spin-polarized scanning tunneling microscopy (SP-STM) in a $PdFe$ bilayer on a $Ir(111)$ substrate at low temperature \cite{ref10}. Leonov et al. conducted both experimental and theoretical studies to investigate the magnetic properties of isolated nanoscale skyrmions in $PdFe/Ir$ bilayers \cite{ref11,ref15}.

From an application standpoint, skyrmion-based technologies offer notable advantages over traditional domain-wall-based concepts, such as racetrack memory. Although both approaches rely on electrically induced motion of magnetic textures, skyrmions can be displaced with substantially lower current densities (on the order of $10^5-10^7$ $A/m^2$) than domain walls (approximately $10^{11}-10^{12}$ $A/m^2$)\cite{ref55,ref56}. This dramatic reduction enables significantly lower power consumption and mitigates Joule heating effects. Moreover, their topological nature enhances resilience against structural imperfections, enabling faster, more reliable information transport at high storage densities. These properties position skyrmions as strong candidates for next-generation, energy-efficient data storage and logic architectures.  \cite{ref12}. Due to these remarkable features, skyrmions have attracted significant interest in recent years, having been studied theoretically and experimentally \cite{ref15,ref55,ref56,ref13,ref14}.

In addition to skyrmions, a variety of other topological formations can also arise in magnetic materials, giving rise to a range of eccentric physical phenomena \cite{ref16,ref46} such as meron and antimeron \cite{ref41}. For example, a skyrmion can be conceptually split into two parts, yielding a meron and an antimeron. Unlike a full skyrmion, a meron covers only half of the spin sphere, reaching to a fractional topological charge of $Q=\pm\frac{1}{2}$. In recent years, numerous theoretical studies have proposed the potential formation of merons and antimerons in chiral magnets exhibiting easy-plane magnetic anisotropy \cite{ref17}. Yu and co-workers provided the first experimental evidence of merons and antimerons in a chiral magnetic film. They successfully stabilized a square lattice composed of alternating merons and antimerons by applying a $20$ mT magnetic field perpendicular to the film at room temperature $(295 K)$ \cite{ref18}. In contrast to the typically observed hexagonal skyrmion lattices, this arrangement reveals an alternating magnetization model that aligns with the theoretical model of bimeron structures \cite{ref19}. In particular, studies have identified three fundamental components for stabilizing the triangular meron–antimeron crystal (MAX): biquadratic interaction, DMI, and easy-plane single-ion anisotropy. Furthermore, by changing the direction of the magnetic field, they show how each magnetic field component affects the triangular $MAX$ \cite{ref20}.

Understanding the magnetic phase behavior in bilayer systems is of great importance for the development of spintronic technologies. Traditionally stabilized by DMI, recent theoretical and experimental studies have shown that alternative mechanisms, such as magnetic frustration and anisotropy, can also support skyrmion phases. Previous studies on easy-axis anisotropy have focused on fundamental magnetic configurations, such as cycloidal structures and skyrmion lattices, and on the stability of individual skyrmions under varying magnetic fields \cite{ref5, ref6, ref11}. Notably, the field-induced instability and collapse of individual skyrmions, as reported in \cite{ref11}, provide a theoretical basis for interpreting experimental observations in systems such as PdFe/Ir(111) bilayers. Banerjee et al. demonstrate that replacing uniaxial anisotropy with two-dimensional easy-plane anisotropy greatly improves the stability of skyrmions under applied magnetic fields \cite{ref21}. Flacke et al. experimentally confirmed this phenomenon by showing that two-dimensional easy-plane anisotropy in magnetic multilayers plays a key role in stabilizing skyrmions \cite{ref22}. In particular, magnetic systems with strong spin–orbit coupling (SOC) naturally exhibit easy-plane, or “compass,” anisotropy, which plays a key role in spin texture formation \cite{ref23, ref24}. Furthermore, researchers have shown that applying an in-plane magnetic field can stabilize skyrmion crystal (SkX) phases in easy-plane magnetic systems, which have traditionally been investigated in relation to the DMI \cite{ref25}. Recent studies have shown that square-lattice skyrmion crystals can be stabilized in centrosymmetric bilayer systems via a staggered DMI. This effect, together with interlayer exchange coupling, can give rise to complex magnetic states and provide novel approaches for creating and manipulating topological patterns in centrosymmetric magnetic systems \cite{ref26,ref27}.

Interlayer interaction occurs between adjacent atomic layers. It can override intralayer interaction, which happens within a single atomic layer. This may result in exotic magnetic phases with nonlinear spin textures \cite{ref49,ref53}. While this single-layer approach predicts topological magnetic phases in certain magnetic moiré systems, it does not fully capture the complexity of the moiré bilayer. In Ref \cite{ref49}, the complete spin dynamics of the bilayer reveal that the emergence of non-trivial interlayer moiré domains can stabilize various spin textures, such as skyrmions, antiskyrmions, and higher-order spin textures, even without conventional sources like  DMI, and dipole interactions. Experiments \cite{ref52} have supported many theoretical predictions of nonlinear moiré magnetic states, especially in $CrI_3$ moiré superlattices. Building on these experimental observations, theoretical studies of moiré magnets have further simplified the moiré bilayer by assuming that magnetic moments in a single layer remain stable \cite{ref50,ref51}.

As can be understood from the brief literature given above, topological spin textures are one of the most active areas of research in recent times, both experimentally and theoretically, due to the revolutionary possibilities promised by technological applications.
However, the mechanisms controlling the transformation and stability of these textures in coupled magnetic layers remain incompletely understood.
Thus, the main aim of this work is to determine the effects of easy-axis and easy-plane anisotropy on the magnetic bilayer system.

For this aim the paper is organized as follows: In Sec. \ref{formulation}, we briefly present the model and formulation. The results and discussions are presented in Sec. \ref{results}, and finally, Sec. \ref{conclusion} contains our conclusions.

\section{Model and Formulation}\label{formulation}

The classical Heisenberg model Hamiltonian for the magnetic bilayer system on a square lattice is given by,
%
%
\begin{equation}\label{eq:1}
\displaystyle
\begin{array}{lcl} \displaystyle
\mathcal{H}=-J_{intra}\sum_{\alpha=\mu,\nu}\sum_{<i,j>}{}{{\vec{S_{i}^\alpha}}\cdot {\vec{S_{j}^\alpha}}}-J_{inter}\sum_{i}{}{{\vec{S_{i}^\mu}}\cdot {\vec{S_{i}^\nu}}}-\sum_{\alpha=\mu,\nu}\sum_{<i,j>} {\vec{D_{ij}^{\alpha}}} ({\vec{S_{i}^\alpha}} \times {\vec{S_{j}^\alpha}}) \\
  \displaystyle
\hspace{2.0cm} -\sum_{\alpha=\mu,\nu}\sum_{i}h_{i}S_i^{z\alpha}-K\sum_{\alpha=\mu,\nu}\sum_{i}(S_i^{z\alpha})^{2},
\end{array}
\end{equation}
where $\alpha=\mu, \nu$ are the layer indices that correspond to the top and bottom layers in the bilayer system. $\vec{S_i}$ denotes the  three-component unit spin vector on the $i$th lattice site. The first sum represents intralayer exchange interactions, whereas the second sum represents interlayer one. Here $<i,j>$ denotes the sum over the nearest neighbors. The layers $\mu$ and $\nu$ are coupled by the inter-layer exchange interaction $J_{inter}$. The Hamiltonian involves ferromagnetic (FM) interaction strengths for both $J_{intra}>0$ and $J_{inter}>0$ as exchange interactions. The third term is the layer-dependent DMI vector and is indicated by $\vec{D_{ij}}=D\vec{r_{ij}}/|\vec{r_{ij}}|$, where $D$ is the magnitude of the DMI vector and $\vec{r_{ij}}$ is the radius vector between two neighboring sites $i$ and $j$. Bloch and N$\acute{e}$el type skyrmions can be obtained depending on the relationship between the DMI vector and the radius vector. If the DMI vector and the radius vector are parallel, Bloch-type skyrmions are observed; if they are perpendicular, N$\acute{e}$el-type skyrmions are observed. We preferred a Bloch-type skyrmion to represent the case in which these vectors are aligned with each other. Skyrmions exhibit swirling motion in the right-handed or left-handed direction, depending on whether the DMI vector is positive or negative, respectively. The parameter $h_{i}$ in the next term is the external magnetic field applied longitudinally at the lattice site $i$. $K$ is the crystal field parameter in the final term, which stands for an easy-axis $K>0$ single-ion anisotropy that tends to align the spin vector in the z direction, and an easy-plane $K<0$ anisotropy that favors the in-plane alignment of the spin vector.

The calculations are performed by using the Monte Carlo simulation
\cite{ref28, ref29} method based on the Metropolis algorithm to study the
scalar chirality map of the magnetic bilayer system model we presented above. We
consider each layer to consist of a two-dimensional square lattice $(L
\times L \times 2)$ within periodic boundary conditions, where $L=50$. After
taking steps up to $5 \times 10^5$ Monte Carlo steps per spin (MCSs) for
equilibrium, the first $10^5$ MCSs have been discarded. Subsequently, the
remaining data (MCSs) were used to perform numerical calculations for
thermalization
averaging.
These
values were
determined
by comparing the
results obtained from simulations that have different number of MCS for various Hamiltonian
parameters. The temperature is gradually reduced to $T=0.001J/k_B$, which is
approximately taken as the ground-state temperature of the system. Here, the
Boltzmann constant $k_B=1$. To simplify the model, we set $J_{intra}
=J_{inter}=J=1$. This means we use scaled (dimensionless) quantities
throughout this work, such as $K/J$, $h/J$, and $D/J$.

To identify the presence of swirling spin textures, we introduce the local chirality $\chi_{i}^{\alpha}$ at the lattice site $i$, at each layer, which is given by,

\begin{equation}\label{eq:2}
\displaystyle
\begin{array}{lcl} \displaystyle
\chi_{i}^{\alpha}=\frac{1}{8\pi}\left[{\vec{S_{i}^\alpha}}\cdot({\vec{S}{i+\hat{x}}^\alpha}\times {\vec{S}{i+\hat{y}}^\alpha})+{\vec{S}{i}^\alpha}\cdot({\vec{S}{i-\hat{x}}^\alpha}\times {\vec{S}_{i-\hat{y}}^\alpha})\right]\end{array}
\end{equation}

If we take the summation of $\chi_{i}^{\alpha}$ over the whole lattice in each layer $\alpha=\mu$ or $\nu$, we get the total scalar chirality, the so-called topological winding number or skyrmion number, as follows:

\begin{equation}\label{eq:3}
\displaystyle
\begin{array}{lcl} \displaystyle
\chi_{T}^{\alpha}=\sum_{i\in\alpha}{\chi_i^{\alpha}}.\end{array}
\end{equation}

We investigate the static spin structure factor in the reciprocal lattice to analyze the magnetic spin Bragg peaks in detail. The perpendicular $S_\bot^\alpha(\vec{q})$ and parallel (to z) $S_{||}^\alpha(\vec{q})$ components of the spin structure factor for each layer are defined by:

\begin{equation}\label{eq:4}
\displaystyle
\begin{array}{lcl} \displaystyle
S_\bot^\alpha(\vec{q})=\frac{1}{N}\bigg\langle \bigg|\sum_{r}{S_{\vec{r}}^{x\alpha}} \cdot e^{-i\vec{q}\cdot \vec{r}}\bigg|^2+\bigg|\sum_{r}{S_{\vec{r}}^{y\alpha}} \cdot e^{-i\vec{q}\cdot \vec{r}}\bigg|^2\bigg\rangle\end{array}
\end{equation}

\begin{equation}\label{eq:5}
\displaystyle
\begin{array}{lcl} \displaystyle
S_{||}^\alpha(\vec{q})=\frac{1}{N}\bigg\langle \bigg|\sum_{r}{S_{\vec{r}}^{z\alpha}} \cdot e^{-i\vec{q}\cdot \vec{r}}\bigg|^2\bigg\rangle\end{array}
\end{equation}
where $<…>$ means the averaged MC configurations and $N=L^2$ is the total number of spins.

\section{Results and Discussion}\label{results}

\begin{figure}[!htbp]
\centering
\includegraphics[width=10cm,angle=0,height=8cm]{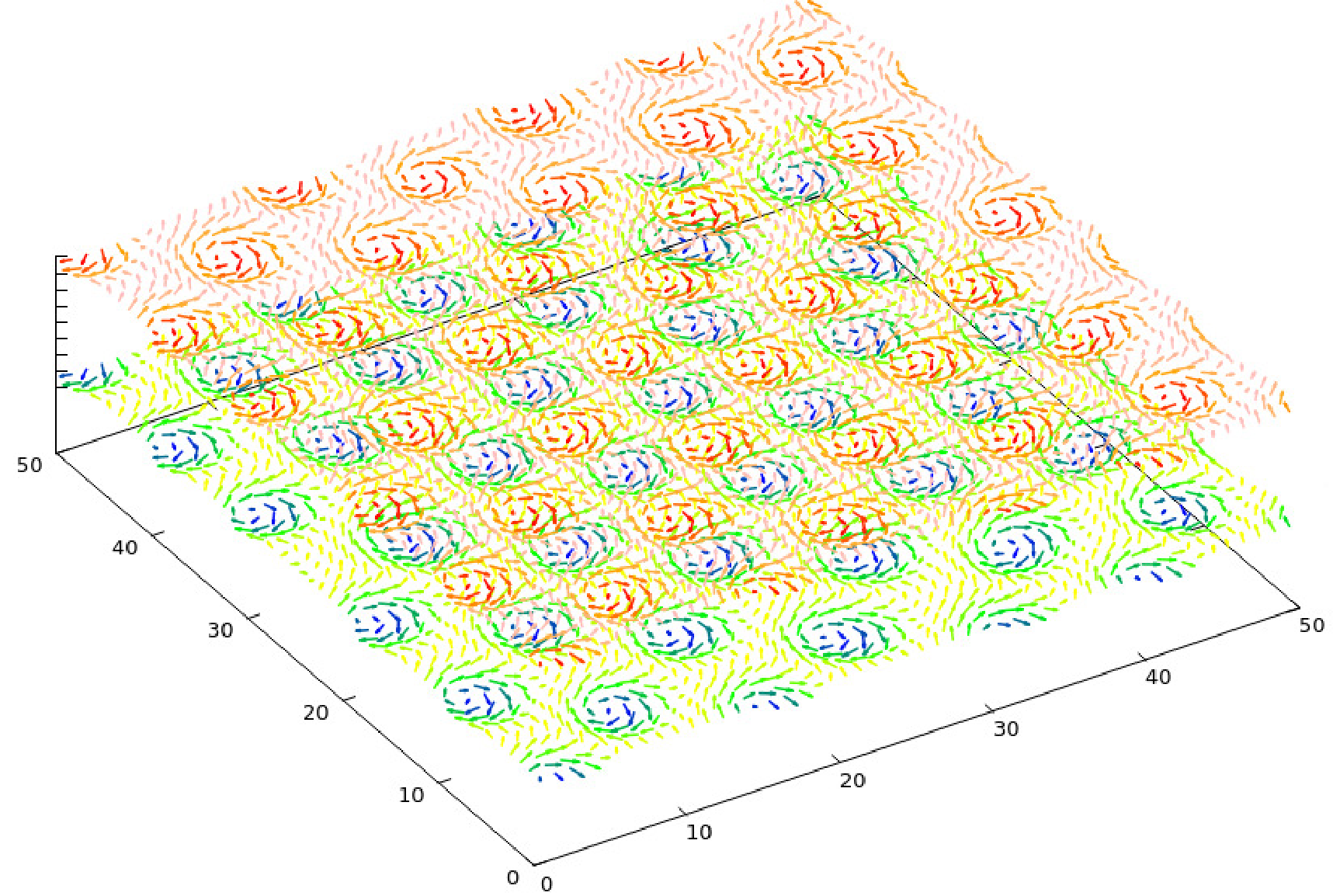}
\includegraphics[width=10cm,angle=0,height=5.0cm]{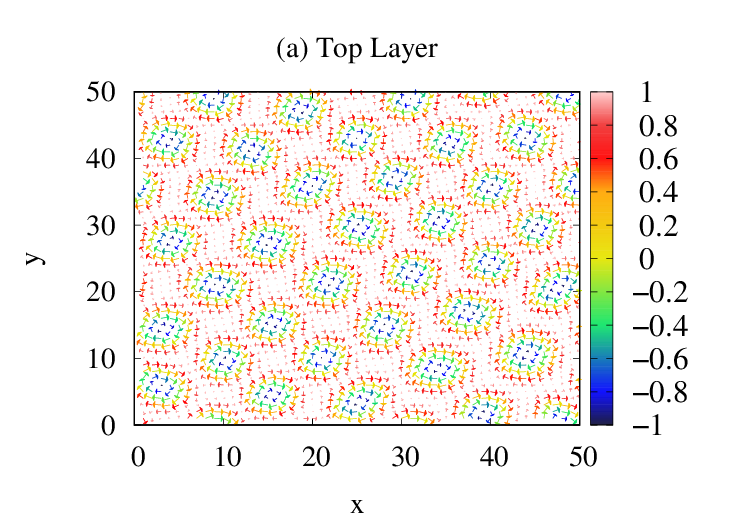}
\includegraphics[width=10cm,angle=0,height=5.0cm]{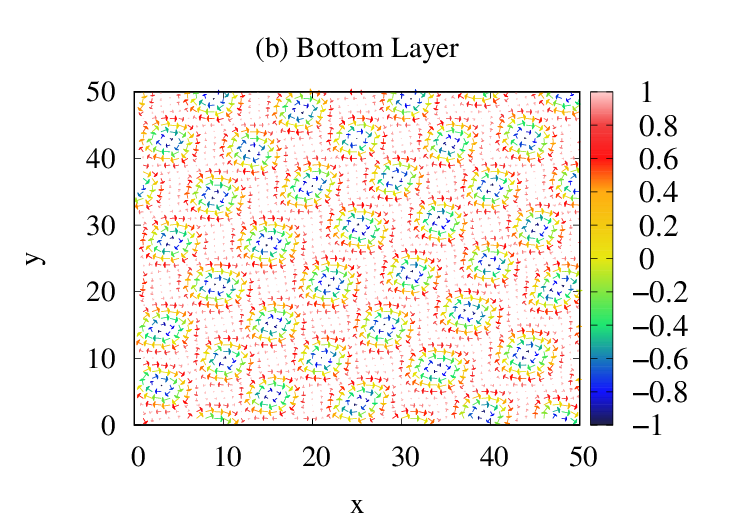}
\caption{Bilayer system  that consists of (a) the top and (b) bottom layer for selected parameters: $D/J = 1$, $h/J = 0.5$, and $K/J = -0.3$.}
\label{fig1}
\end{figure}
We designed the magnetic bilayer system as shown in Fig. \ref{fig1}. Our
system consists of two layers, (a) the top layer and (b) the bottom layer,
each a square lattice. In the Fig. \ref{fig1}, SkX
phases emerge in both layers for the following values: $D/J=1$, $h/J=0.5$,
and $K/J=-0.3$. Since the interaction between the layers is ferromagnetic
$J_{inter} \neq 0$, the system behaves synchronously, then we prefer to
display the results of one monolayer.
However, the bilayer system is physically relevant because it introduces stabilization mechanisms that are absent in a monolayer. The interlayer exchange interaction  term modifies the topological energy landscape. This leads to a broader stability region of the observed topological spin textures. Another reason for framing the problem as a system consisting of two identical layers with ferromagnetic interaction is that subsequent studies
will investigate the effects of
changes in values such as
interlayer interaction and intralayer interaction on spin textures.

We performed simulations for several values of the Dzyaloshinskii–Moriya interaction parameter.  Although the quantitative boundaries of the different spin textures  shift with increasing $D/J$, the overall topology of the scalar chirality map and the sequence of magnetic textures remain qualitatively unchanged. For clarity and to avoid redundancy, we present the detailed analysis for a representative value  $D/J=1.$

To investigate the spin textures in the system, we construct the scalar chirality map in  $(K/J-h/J)$ plane. After giving this map, we will provide spin snapshots, local chirality maps, and structure factors for selected values of the Hamiltonian parameters. Spin snapshots are shown as vector plots of the spins in the $xy$ plane; the vector colors are determined by the $z$ components of the spins, as shown in the colorbar for each snapshot. Similarly, local chirality maps are shown in $xy$ plane where one layer of the bilayer resides. Lastly, the spin structure factors are depicted in the $q_x,q_y$ plane.

After presenting the scalar chirality map, we first examine the effect of anisotropy in the zero-field case, $h/J=0$ case. After this, we investigate the effect of the magnetic field on the spin textures by getting some representative spin snapshots for the isotropic case ($K/J=0.0$), easy axis anisotropy case ($K/J=1.0$), easy plane anisotropy case ($K/J=-1.2$), and for choosing magnetic field values.

\subsection{Ground-State scalar chirality map of Magnetic Bilayer Nanosystems}

\begin{figure}[htp]
\centering
\includegraphics[width=14cm,height=8cm]{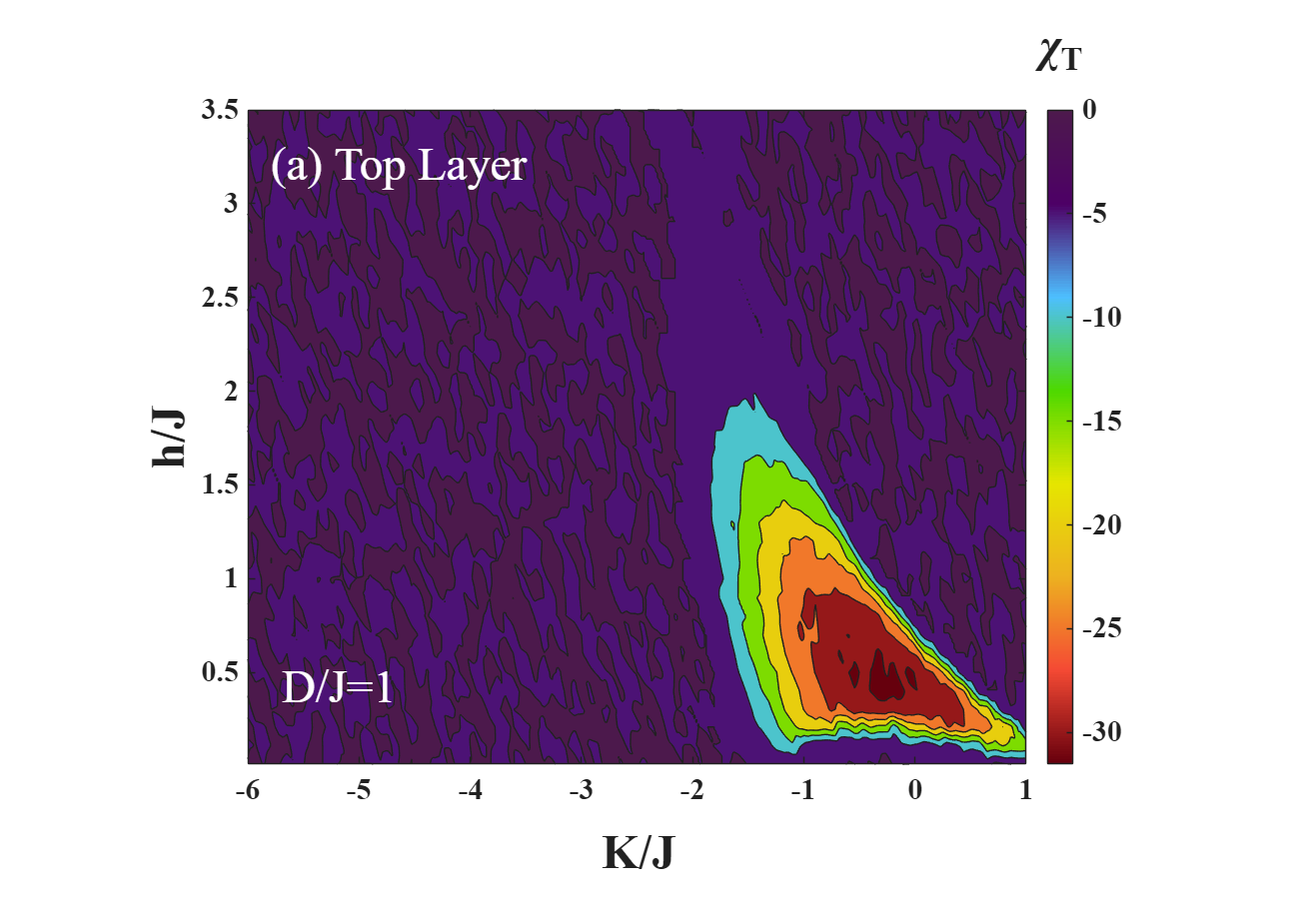} 
\caption{The density plot of the total scalar chirality $\chi_{T}$ as a function of $(K/J-h/J)$ plane for one layer for $D/J=1$.}
\label{fig2}
\end{figure}

Since scalar chirality identifies topological phases, we examine the ground-state  $(T=0.001 J/k_B)$ scalar chirality map to explore these phases, given in Fig. \ref{fig2}. The color bar represents the total scalar chirality in the layer.  A Bloch-type skyrmion with a negative center has topological charge $-1$, which contributes negatively per unit sphere to the total scalar chirality. Thus, we obtain negative values in the color bar. Contour lines are closed curves formed by connecting points of the same chirality, starting from the initial level of the contour. We can address the area with the largest total chirality in scalar terms, the red region, as the region where skyrmions are most abundant. This map is obtained by calculating the scalar chirality from Eq. \ref{eq:3} by scanning the Hamiltonian parameter range of $K/J$ and $h/J$ by steps $0.2$.

The scalar chirality map shown in Fig. \ref{fig2} provides a global overview of the topological textures that emerge in the
$(K/J,h/J)$ parameter space. Since the
total scalar chirality reflects the
density of topological defects in
the system, regions with large magnitude
of chirality indicates the presence
of skyrmion-rich configurations. In contrast, regions where the total
chirality approaches zero may correspond either to topologically trivial
states or to textures composed of meron–antimeron pairs whose local
topological charges cancel each other. Therefore, the chirality map alone is not sufficient to uniquely determine the underlying spin configuration. To identify the corresponding magnetic textures, we analyze representative points in the parameter space by examining real-space spin configurations, local chirality distributions, and spin structure factors. These complementary diagnostics allow us to distinguish between labyrinth structures, skyrmion lattices, and MAX phase.

We are starting at the horizontal  line at $h/J=0$ in this map, i.e., the dark blue region at the bottom in Fig. \ref{fig2}.

\subsection{ Zero magnetic field case}

\begin{figure}[htp]
\centering
\includegraphics[width=7cm,height=7cm]{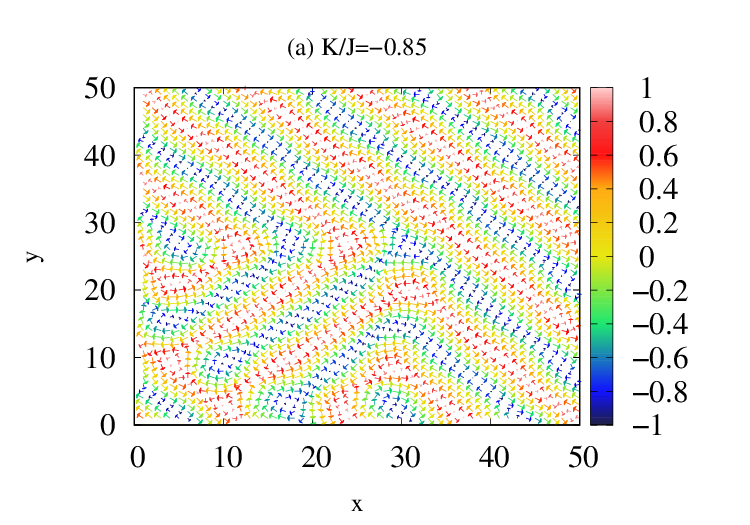}
\includegraphics[width=7cm,height=7cm]{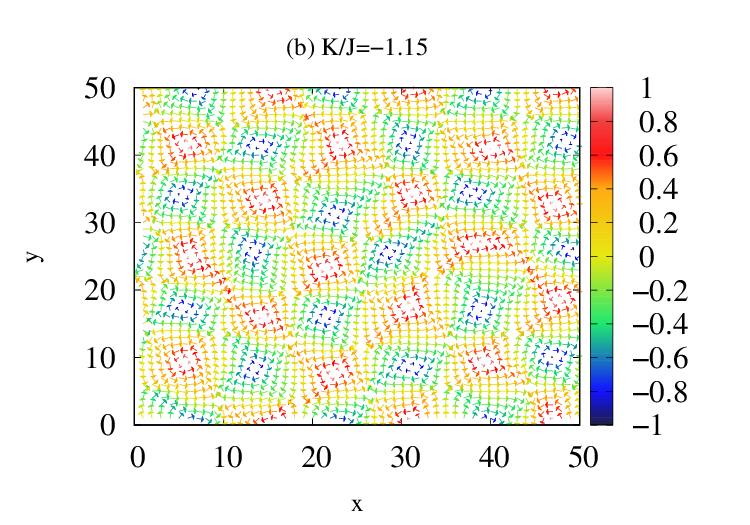}
\includegraphics[width=7cm,height=6cm]{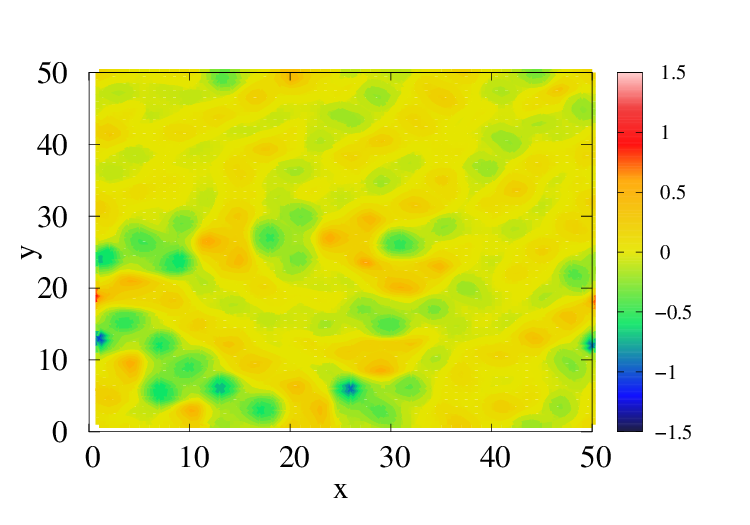}
\includegraphics[width=7cm,height=6cm]{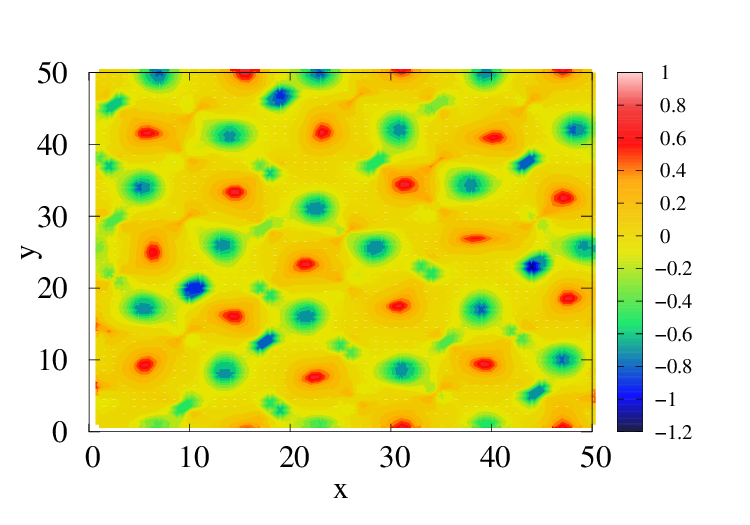}
\includegraphics[width=7cm,height=9cm]{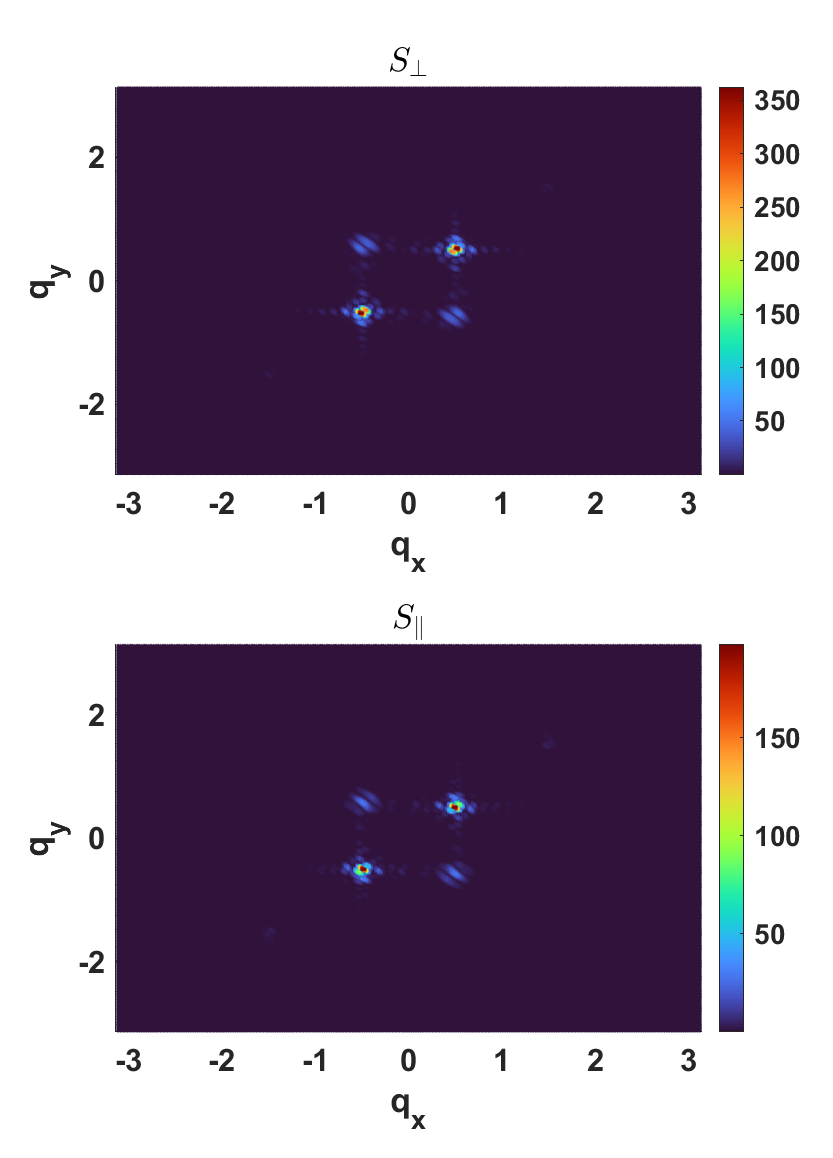}
\includegraphics[width=7cm,height=9cm]{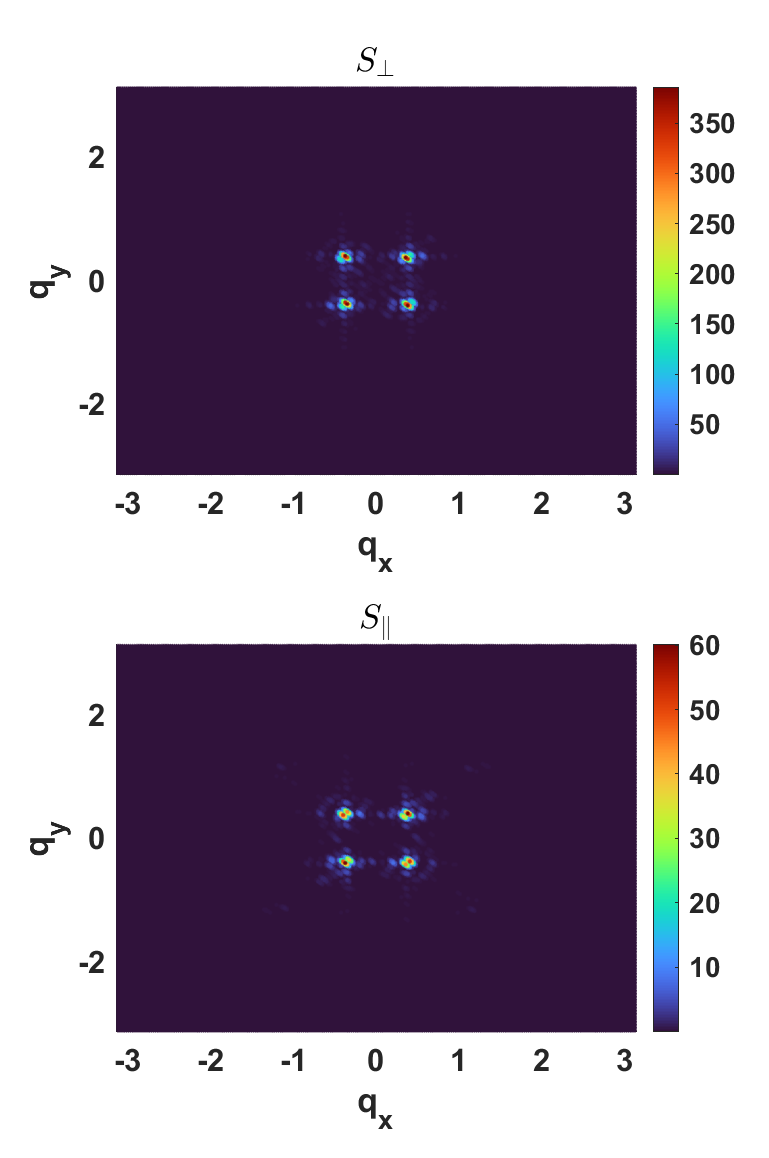}
\caption{The real-space spin configurations (top panel) and real-space local chirality (middle panel) corresponding to that configuration $D/J=1$ without magnetic field $h/J=0$ for selected values of anisotropy (a) $K/J=-0.85$, (b) $K/J=-1.15$, respectively. The perpendicular $S_\bot(\vec{q})$ and parallel $S_{||}(\vec{q})$ components of the static structure factor with the corresponding structure are shown in the two rows below panel.}
\label{fig3}
\end{figure}
\begin{figure}[htp]
\centering
\includegraphics[width=7cm,height=7cm]{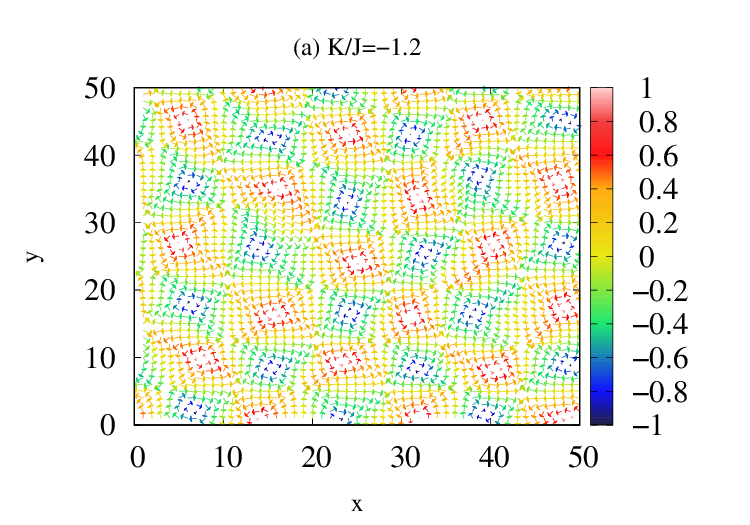}
\includegraphics[width=7cm,height=7cm]{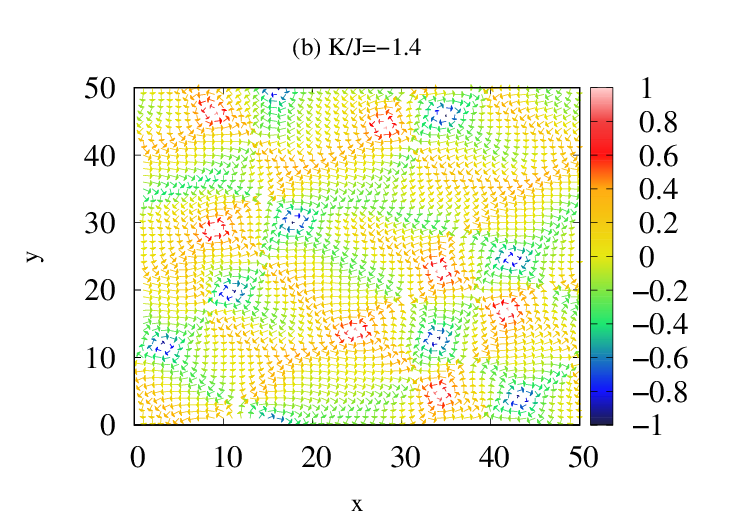}

\includegraphics[width=7cm,height=6cm]{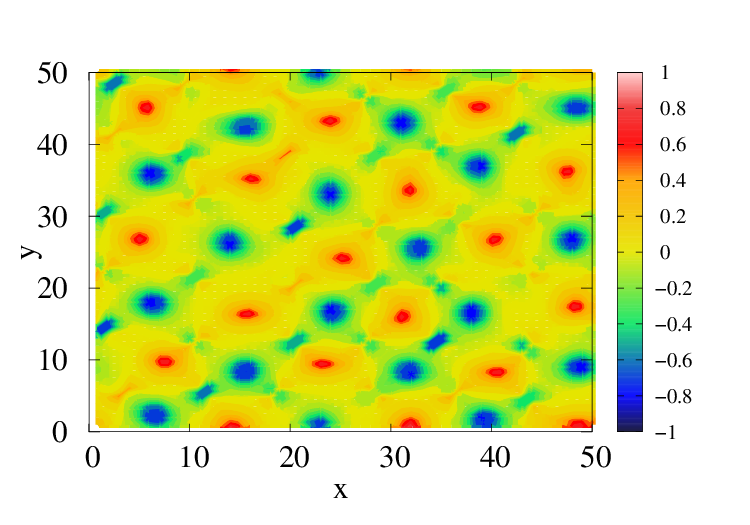}
\includegraphics[width=7cm,height=6cm]{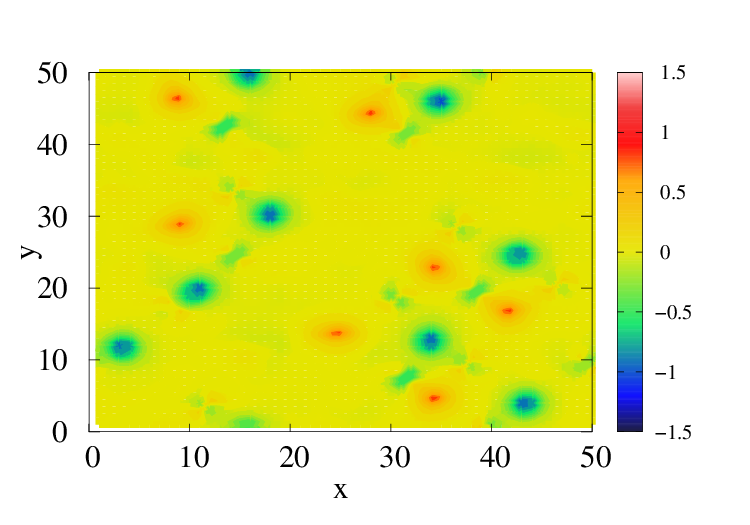}

\includegraphics[width=7cm,height=9cm]{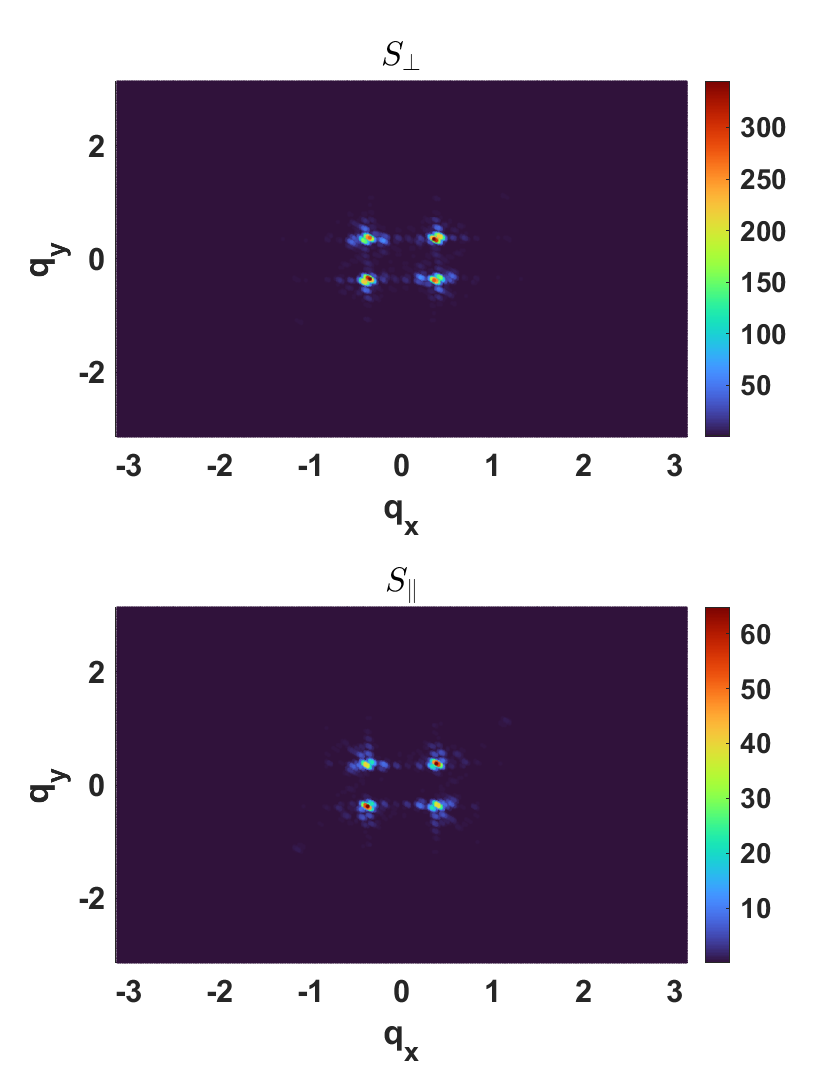}
\includegraphics[width=7cm,height=9cm]{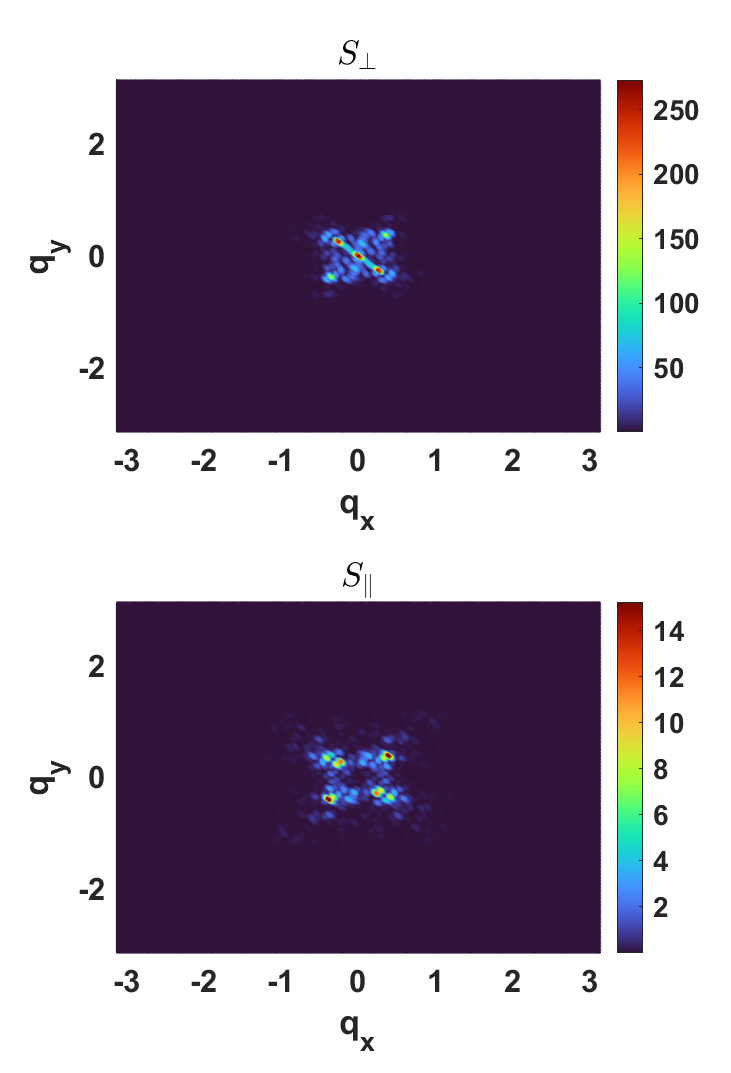}
\caption{The real-space spin configurations (top panel) and real-space local chirality (middle panel) corresponding to that configuration $D/J=1$ without magnetic field $h/J=0$ for selected values of anisotropy (a) $K/J=-1.2$, (b) $K/J=-1.4$, respectively. The perpendicular $S_\bot(\vec{q})$ and parallel $S_{||}(\vec{q})$ components of the static structure factor with the corresponding structure are shown in the two rows below panel.}
\label{fig4}
\end{figure}

The effect of easy-plane anisotropy in the absence of a magnetic field is shown in Figs. \ref{fig3} and \ref{fig4}. We depict the real-space spin configurations in the top row and local chirality maps in the middle row. The bottom rows contain spin structure factors. As seen in Figs. \ref{fig3} and \ref{fig4}, rising easy-plane anisotropy energy transforms the spin-spiral/labyrinth texture into a square lattice of merons and antimerons (the MAX phase, compare Figs. \ref{fig3} (b) with (a)). The structure contains several core-down (blue) meron $\chi_{i}=-1/2$ and core-up (red) antimeron $\chi_{i}=1/2$ sequences (see Figs. \ref{fig3} (b) and \ref{fig4} (a)).

While the total chirality map in Fig. \ref{fig2} has almost zero value at $(K/J,h/J)=(-1.15,0.0)$ or $(-1.2,0.0)$, as seen in Figs. \ref{fig3} (b) and \ref{fig4} (a) local chiralities have ordered structure.
This shows that the total chirality map alone is not sufficient to identify the underlying topological spin texture.

This state is also characterized by the double-$Q$ ($2Q$) Bragg peaks with equal intensities in both of the perpendicular $S_\bot(\vec{q})$ and parallel $S_{||}(\vec{q})$ components of the static structure factor, as shown in the bottom lines of Figs. \ref{fig3} (b) and \ref{fig4} (a), while the intensities are unequal for the spin-spiral/labyrinth spin texture (see bottom rows  of Fig. \ref{fig3} (a)).

If we further increase the strength of the easy plane anisotropy (i.e., increasing the magnitude of $K/J$ in the negative direction) spins will try to orient themselves in the $xy$-plane. There will be a significant decrease in the $z$ orientation of the spins. This manifests itself in the spin texture as a smaller number of merons and antimerons as seen in Fig. \ref{fig4} (b).

In the static structure factors, we can see that the $S_\bot(\vec{q})$ Bragg peaks increase with the center peak, and the $S_{||}(\vec{q})$ component decreases (see bottom rows of Fig. \ref{fig4} (b)). If we increase the easy-plane anisotropy sufficiently, components of the structure factor transform into $1Q$ state and begin to concentrate in the central peak, while weak peaks in the $z$-component decrease visibly.

\subsection{The absence of the single-ion anisotropy: Isotropic case}

\begin{figure}[htp]
\centering
\includegraphics[width=7.4cm,height=6cm]{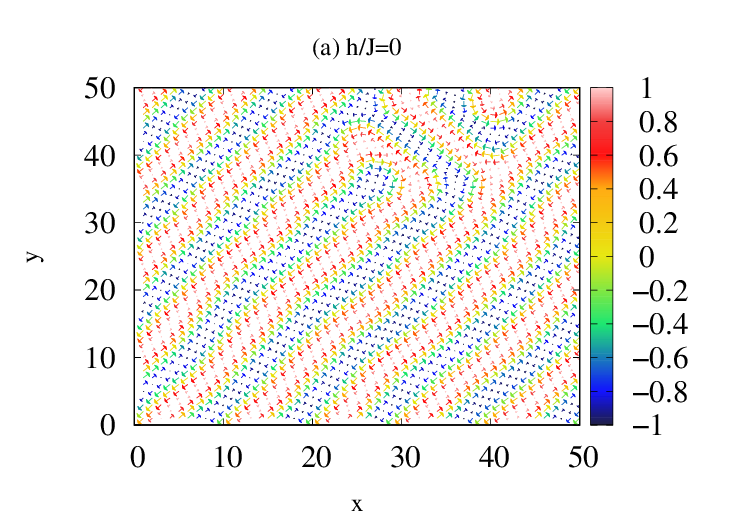}
\includegraphics[width=7.4cm,height=6cm]{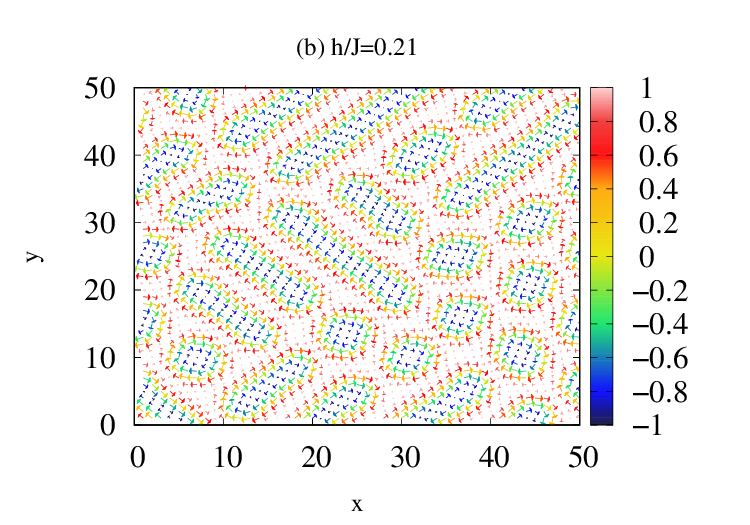}
\includegraphics[width=7cm,height=6cm]{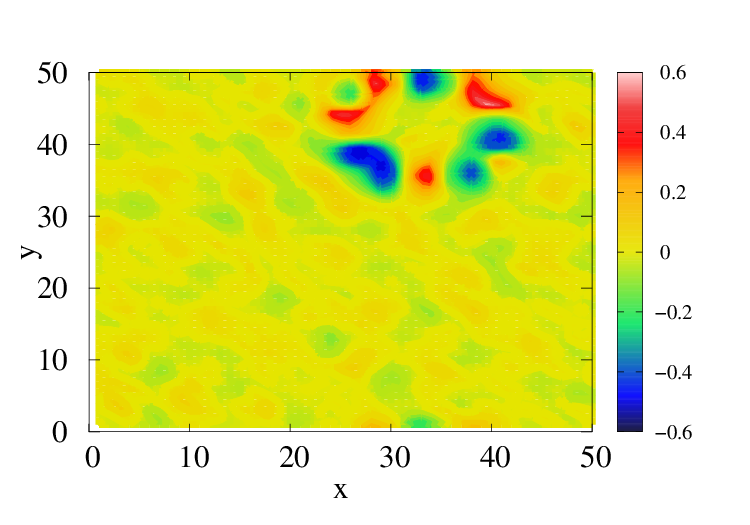}
\includegraphics[width=7cm,height=6cm]{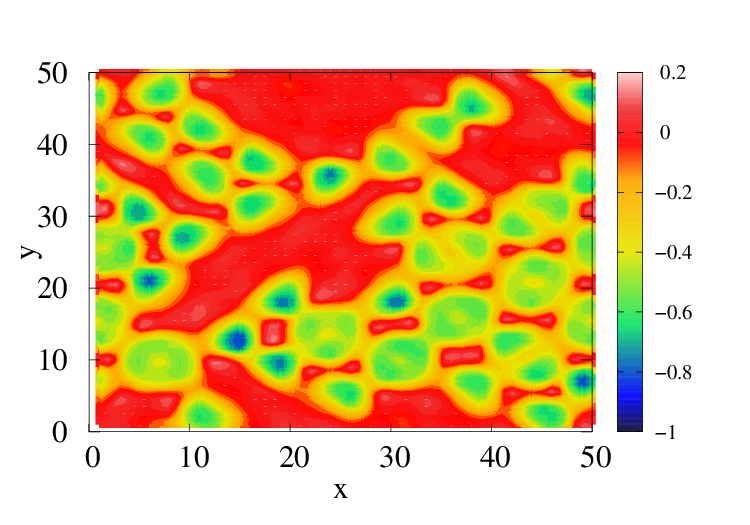}
\includegraphics[width=7.4cm,height=6cm]{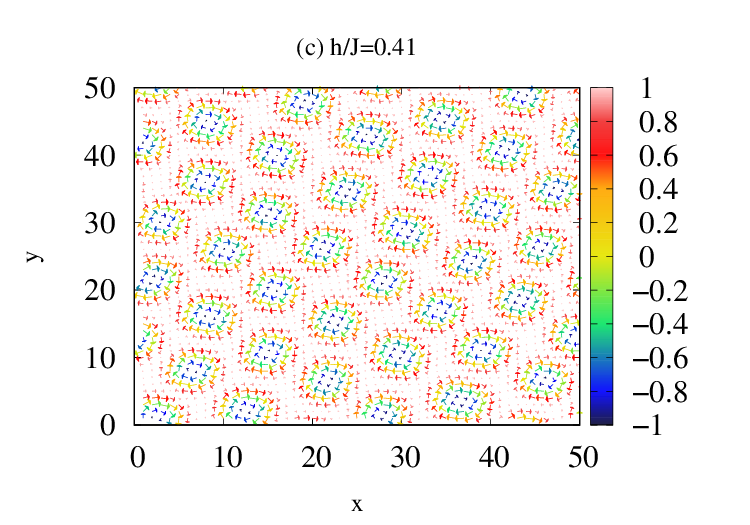}
\includegraphics[width=7.4cm,height=6cm]{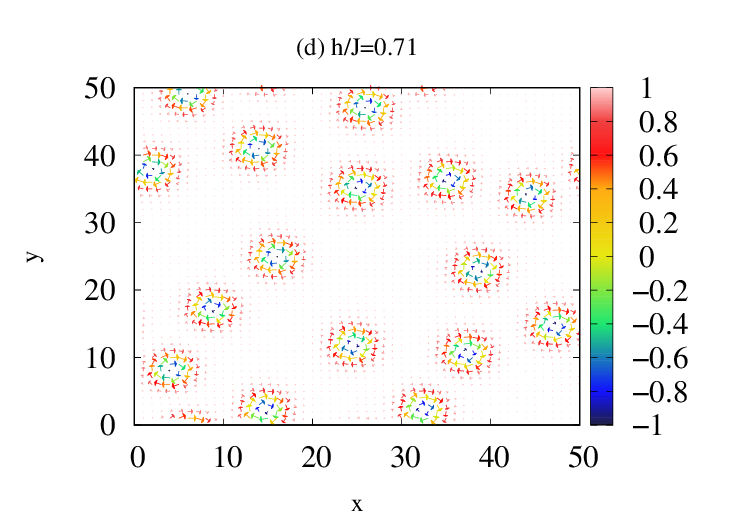}
\includegraphics[width=7cm,height=6cm]{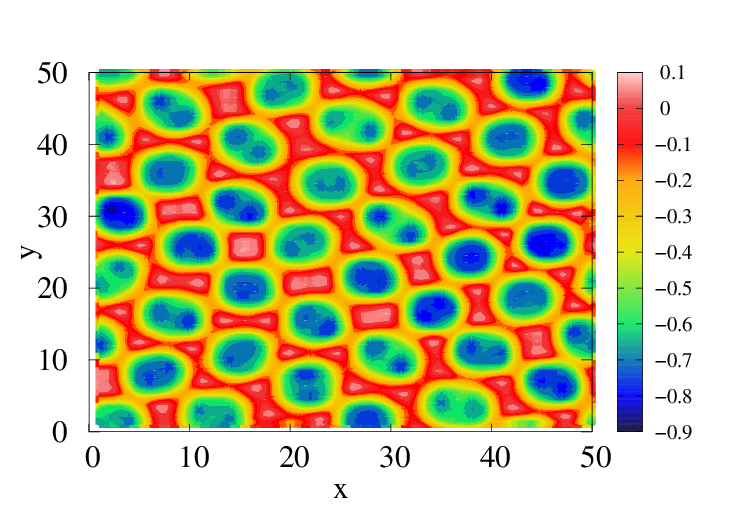}
\includegraphics[width=7cm,height=6cm]{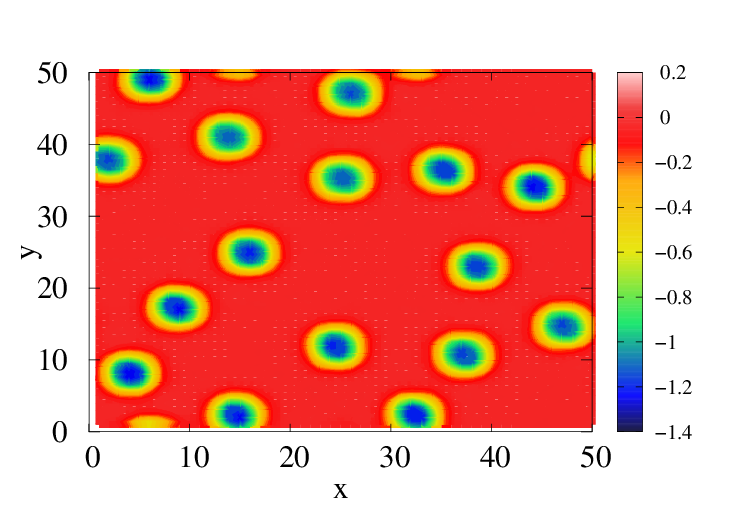}
\caption{Real-space spin configurations (top panel) and real-space local chirality (bottom panel) corresponding to that configuration. Snapshots for $K/J=0$ and $D/J=1$ for selected magnetic field values (a) $h/J=0.0$, (b) $h/J=0.21$, (c) $h/J=0.41$, (d) $h/J=0.71$, respectively.}
\label{fig5}
\end{figure}

In the case of  $K/J=0$, the evolution of the spin textures with magnetic field-induced  along the $z$-direction can be seen in Fig. \ref{fig5}.
Real space spin configurations at zero magnetic field are presented in Fig. \ref{fig5} (a). Almost zero local chirality in ($x,y$) plane and opposite blue and red regions, which cancel each other, is consistent  with the total chirality given in Fig. \ref{fig2} for $(K/J,h/J)=(0.0,0.0)$.
As seen in Fig. \ref{fig5} (b), when we apply the magnetic field to the bilayer, we observe that the labyrinth phase transforms into rod-type domains, and skyrmions are observed together. The system has exited the labyrinthine regime but has not yet fully entered the SkX, indicating an intermediate metastable phase.
This state can  be described as a helical-skyrmion mixed texture.

Researchers explored asymmetric $(Pt/Co/X)$ multilayers with X = Ta or W in ref. \cite{ref57}. Magnetic force microscopy (MFM) captured the W-based sample’s striking shift from rare rod-like domains to skyrmions, while the Ta-based structure transformed its intricate labyrinth domains into skyrmions. Fig. \ref{fig5} (b) corresponds to the intermediate metastable regime where rod-like domains coexist with skyrmions, which is in agreement with the experimental MFM observations reported in \cite{ref57}. It evolved into the SkX phase at $h/J=0.41$ as obtained in Fig. \ref{fig5} (c).

We observed that by increasing the magnetic field to $h/J=0.71$, the spin orientations are forced to move towards out-of-plane, resulting in the coexistence phase of ferromagnetic and skyrmion $(FM+SkX)$ spin texture as seen in Fig. \ref{fig5} (d). The skyrmion patterns begin to distort and disappear when the field increases.

\subsection{The presence of the single-ion anisotropy}
The presence of the single-ion anisotropy can be investigated in two distinct cases: easy-axis anisotropy ($K/J>0$) and easy-plane anisotropy ($K/J<0$)  cases. While in the first case the spins tend to align in the $z$ direction, in-plane alignment is favored in the second case. In both cases, we investigate the effect of the changing field on the spin textures by displaying the representative results for chosen magnetic field values.

\subsubsection{ The easy-axis anisotropy case}

\begin{figure}[htp]
\centering
\includegraphics[width=7cm,height=6cm]{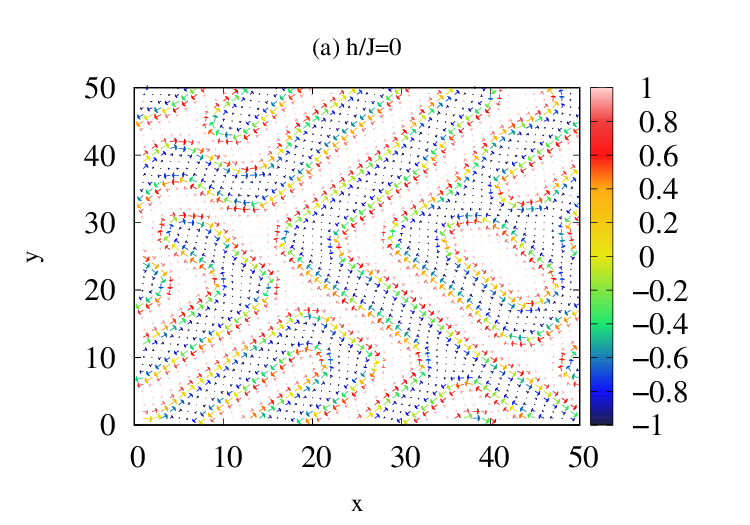}
\includegraphics[width=7cm,height=6cm]{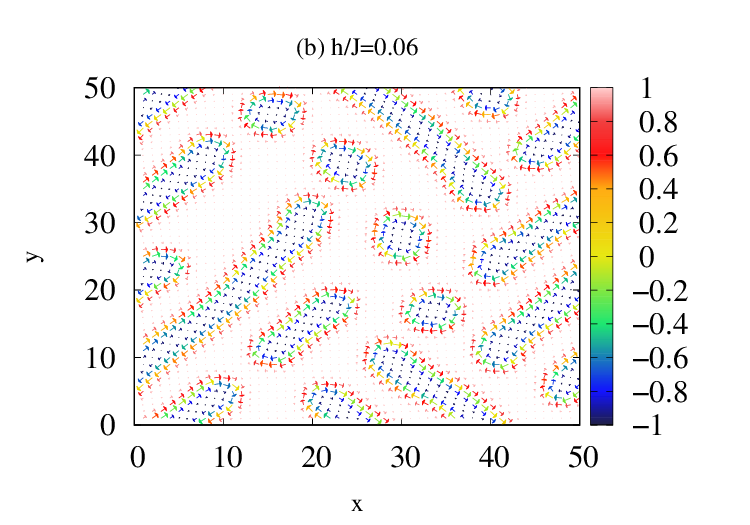}
\includegraphics[width=7cm,height=6cm]{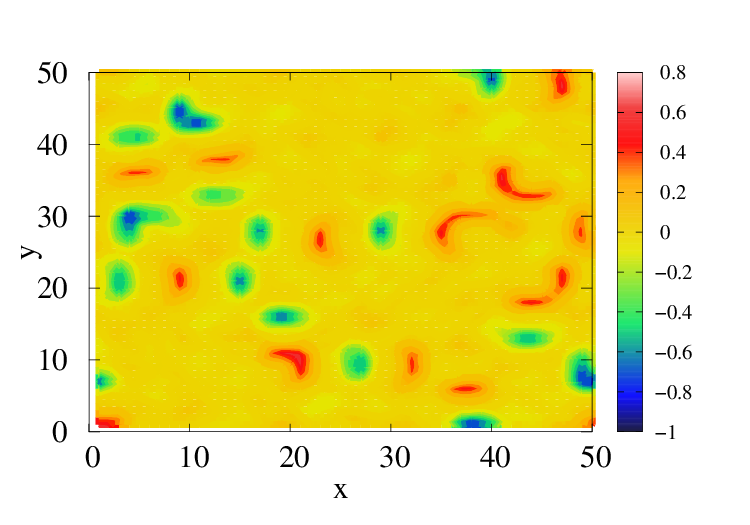}
\includegraphics[width=7cm,height=6cm]{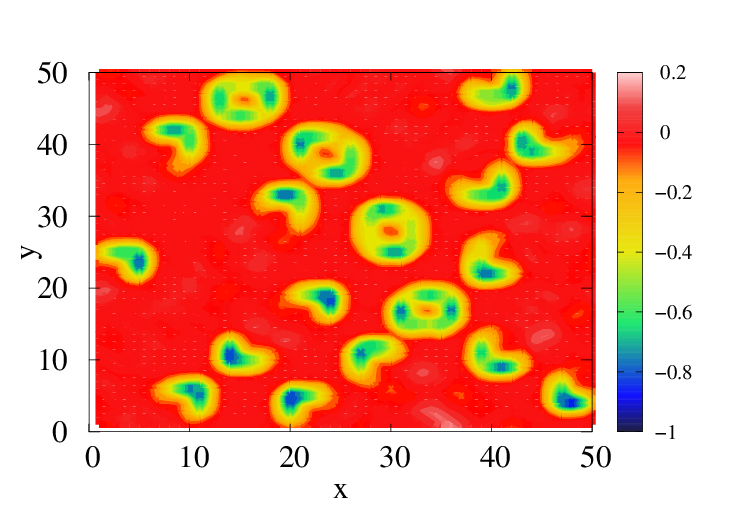}
\includegraphics[width=7cm,height=6cm]{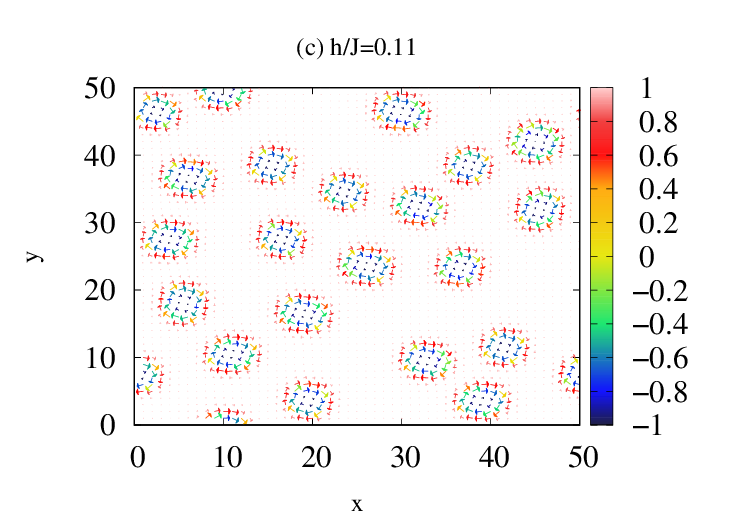}
\includegraphics[width=7cm,height=6cm]{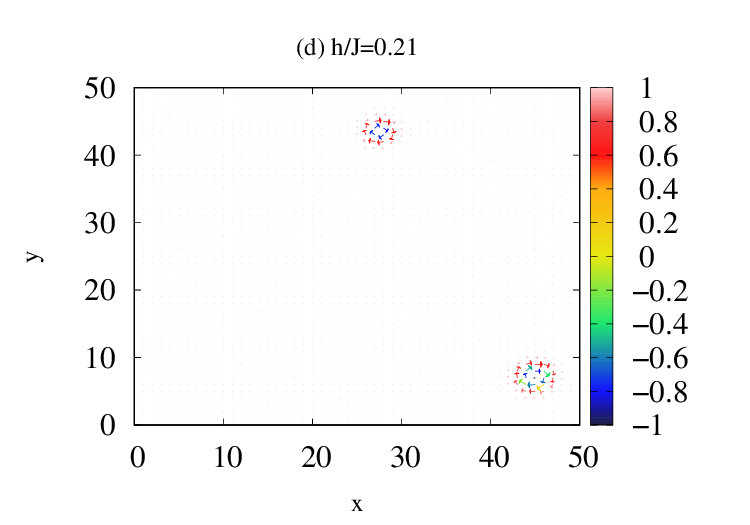}
\includegraphics[width=7cm,height=6cm]{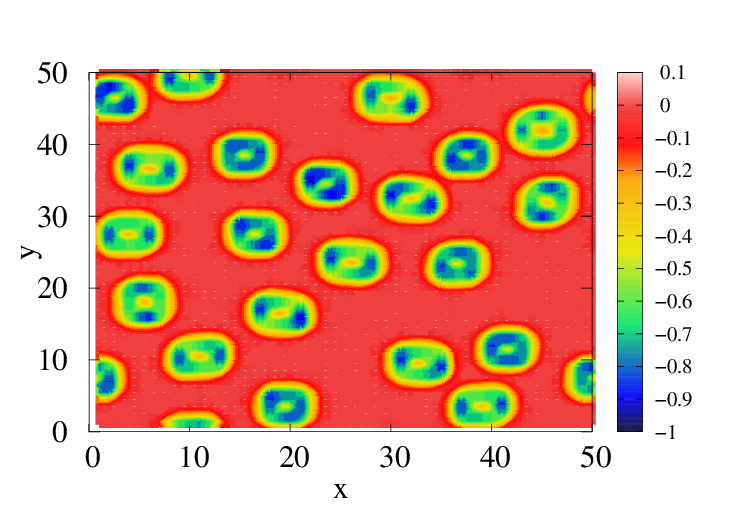}
\includegraphics[width=7cm,height=6cm]{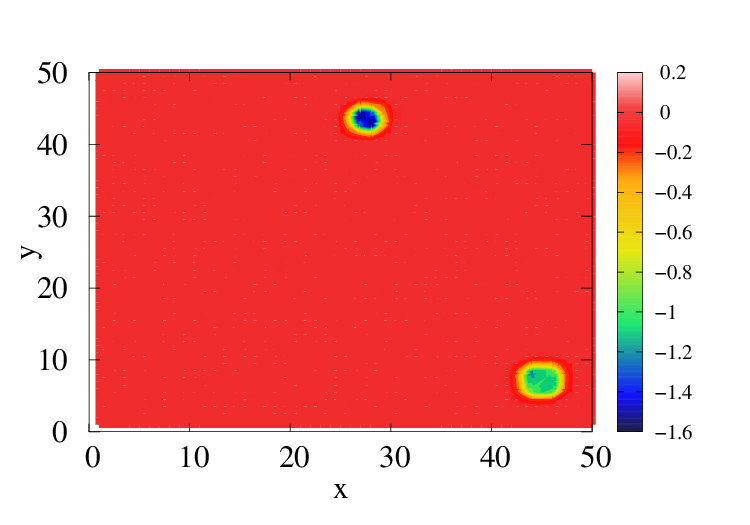}
\caption{Real-space spin configurations (top panel) and real-space local chirality (bottom panel) corresponding to that configuration. Snapshots for $K/J=1$ (easy-axis) and $D/J=1$ for selected (a) $h/J=0$, (b) $h/J=0.06$, (c) $h/J=0.11$, (d) $h/J=0.21$ respectively.}
\label{fig6}
\end{figure}

In this case, we choose anisotropy as $K/J=1$. The system exhibits a labyrinth of domains with more corrugated shapes at $h/J=0.0$ as seen in Fig. \ref{fig6} (a). Even with a minimal magnetic field $h/J=0.06$, that induces  the labyrinth phase is broken into small and separate pieces, skyrmions and stripes are formed in Fig. \ref{fig6} (b).  When we set the magnetic field $h = 0.11$  (Fig. \ref{fig6} (c))   and $h = 0.21$  (Fig. \ref{fig6} (d)), the mixed domains were replaced by ferromagnetic and skyrmion phases $(FM+SkX)$. We observe that the number of skyrmions decreases as the magnetic field strength, $h/J$, increases (compare Fig. \ref{fig6} (d) with (c)). If we further increase the magnetic field, the field-polarized phase takes place.

\subsubsection{The easy-plane anisotropy case }

\begin{figure}[htp]
\centering
\includegraphics[width=7cm,height=7cm]{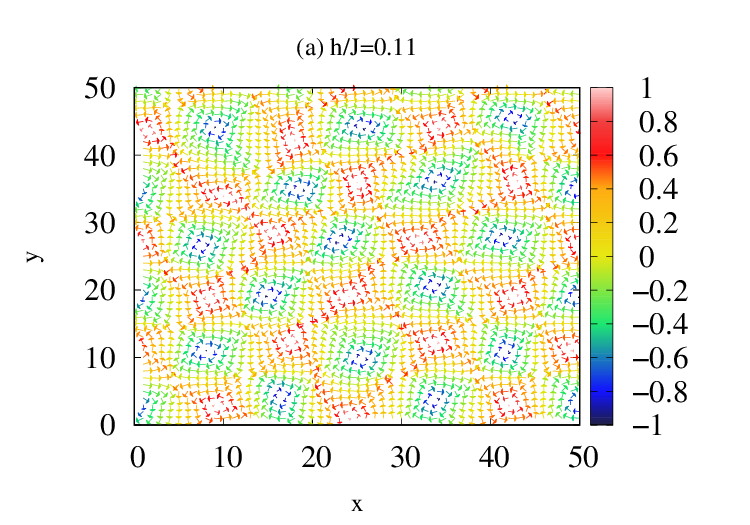}
\includegraphics[width=7cm,height=7cm]{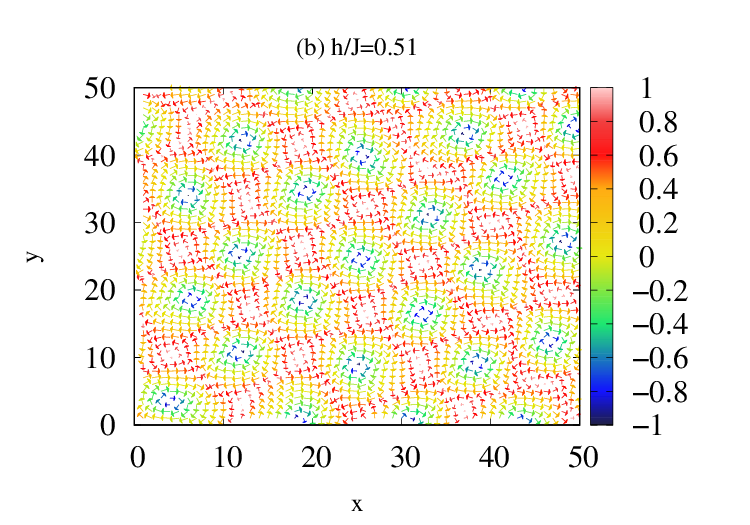}
\includegraphics[width=7cm,height=6cm]{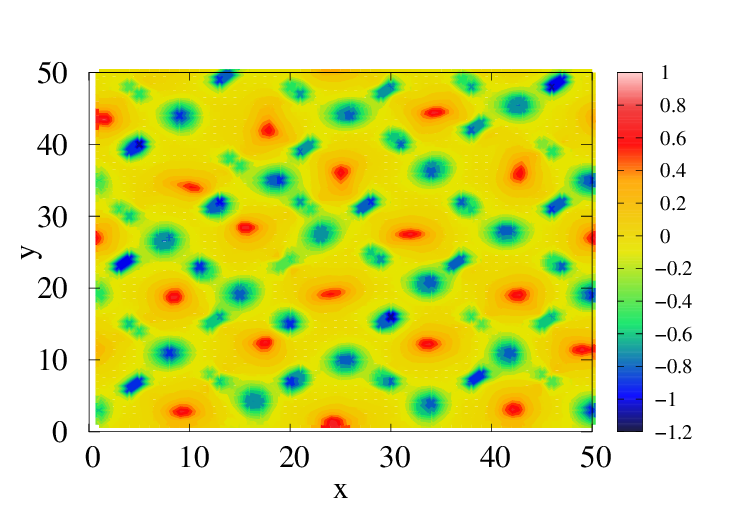}
\includegraphics[width=7cm,height=6cm]{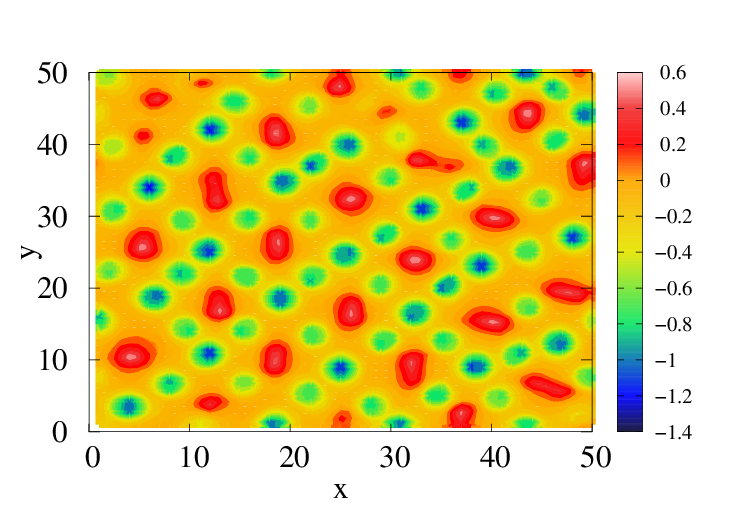}
\includegraphics[width=7cm,height=9cm]{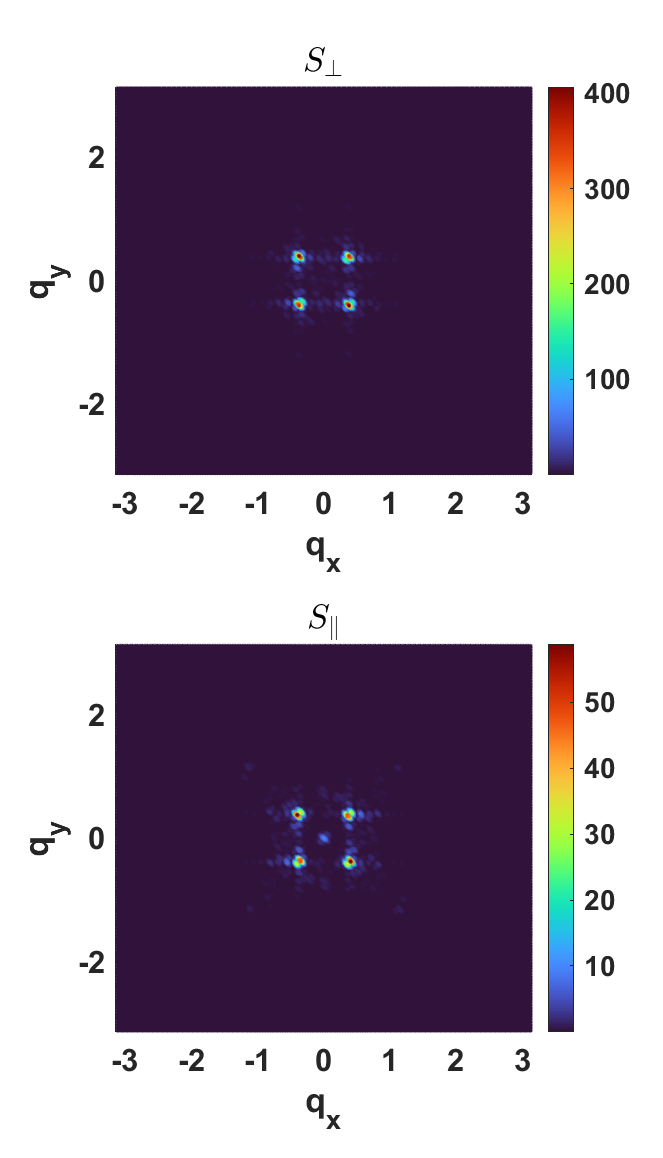}
\includegraphics[width=7cm,height=9cm]{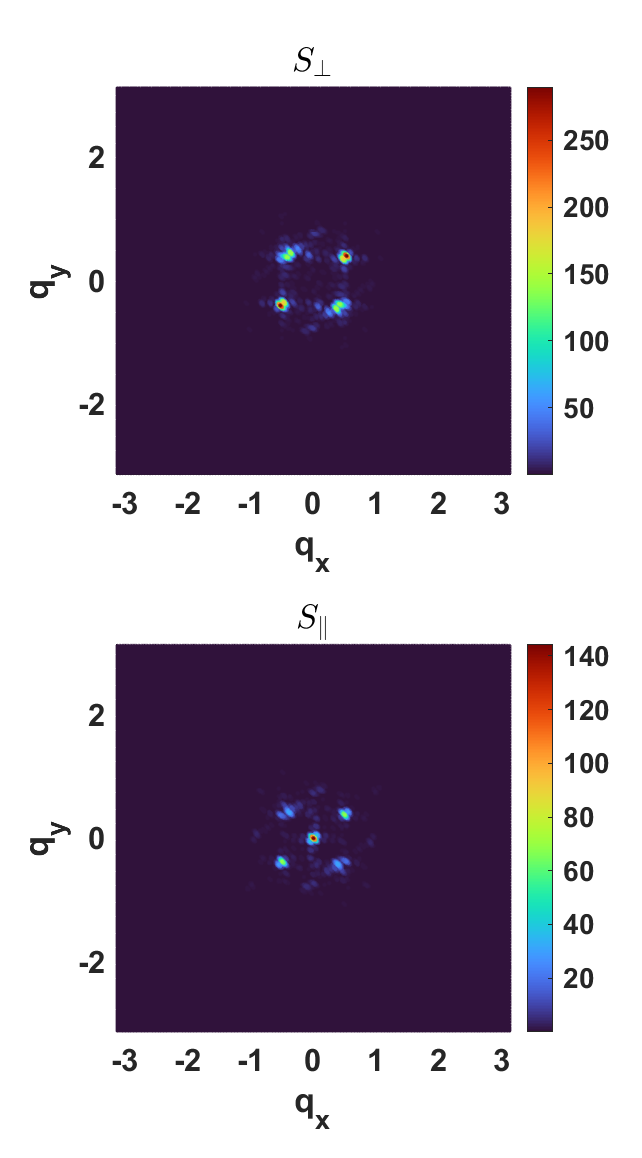}
\caption{The real-space spin configurations for $K/J=-1.2$ and $D/J=1$ for selected (a) $h/J=0.11$, (b) $h/J=0.51$ from left to the right. The middle panel displays the real-space local chirality map corresponding to that configuration. The perpendicular $S_\bot(\vec{q})$ and parallel $S_{||}(\vec{q})$ components of the static structure factor with the corresponding structure are shown in the bottom panel.}
\label{fig7}
\end{figure}

\begin{figure}[htp]
\centering
\includegraphics[width=7cm,height=7cm]{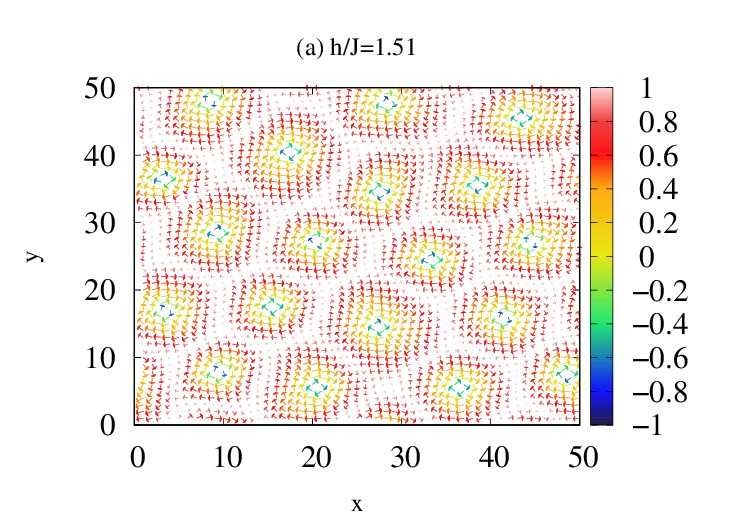}
\includegraphics[width=7cm,height=7cm]{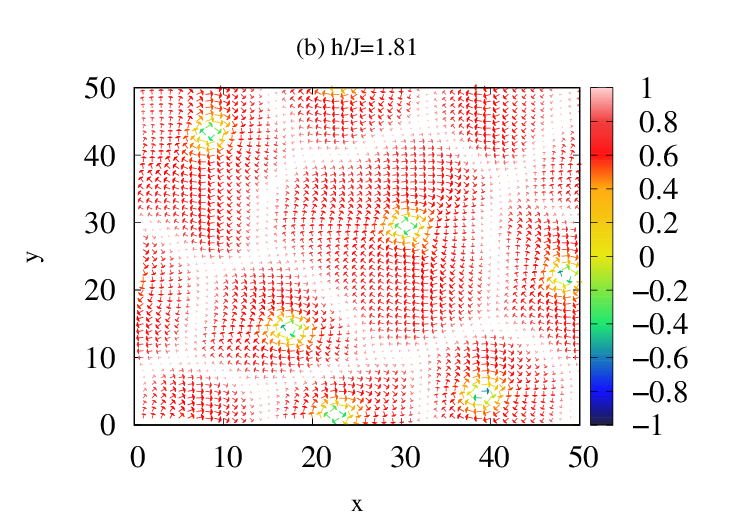}
\includegraphics[width=7cm,height=6cm]{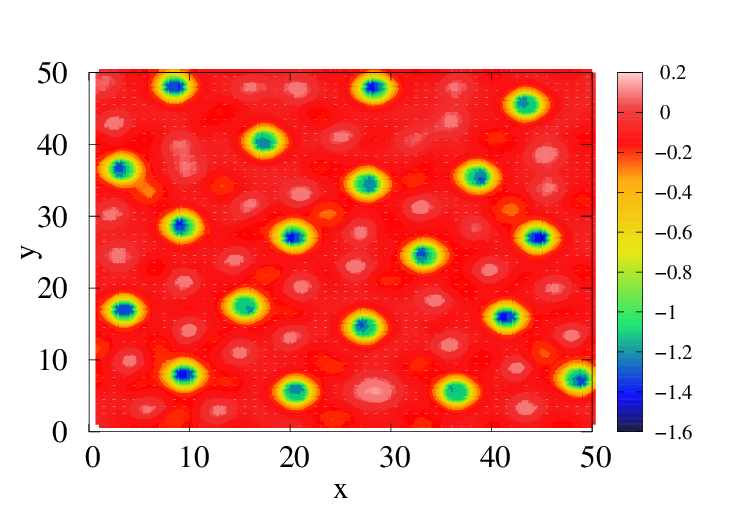}
\includegraphics[width=7cm,height=6cm]{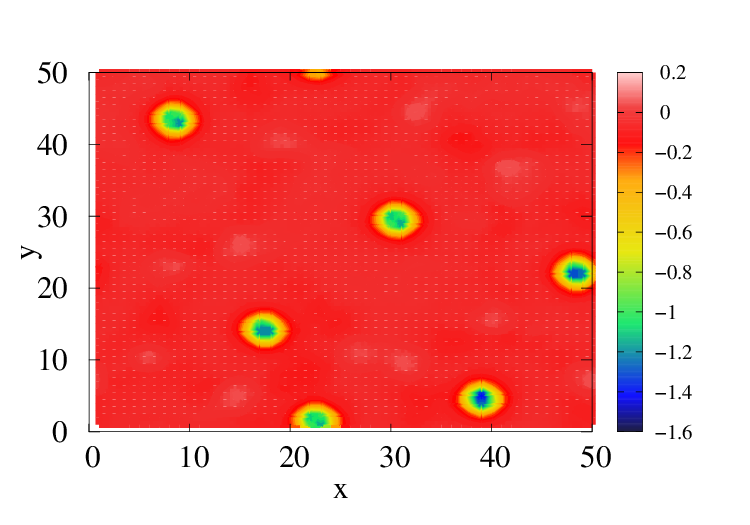}
\includegraphics[width=7cm,height=9cm]{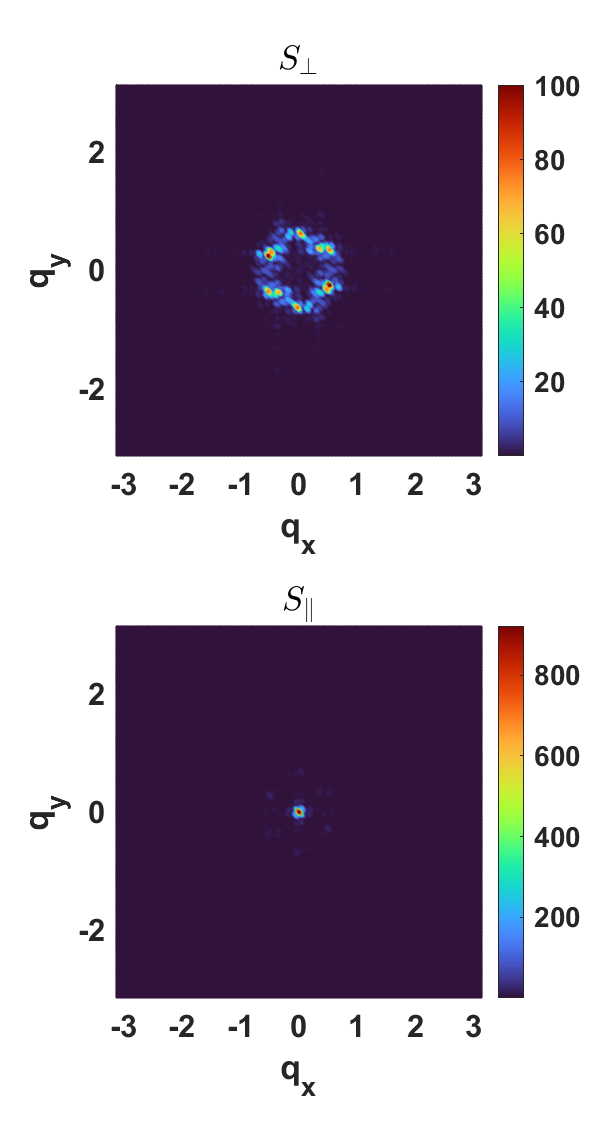}
\includegraphics[width=7cm,height=9cm]{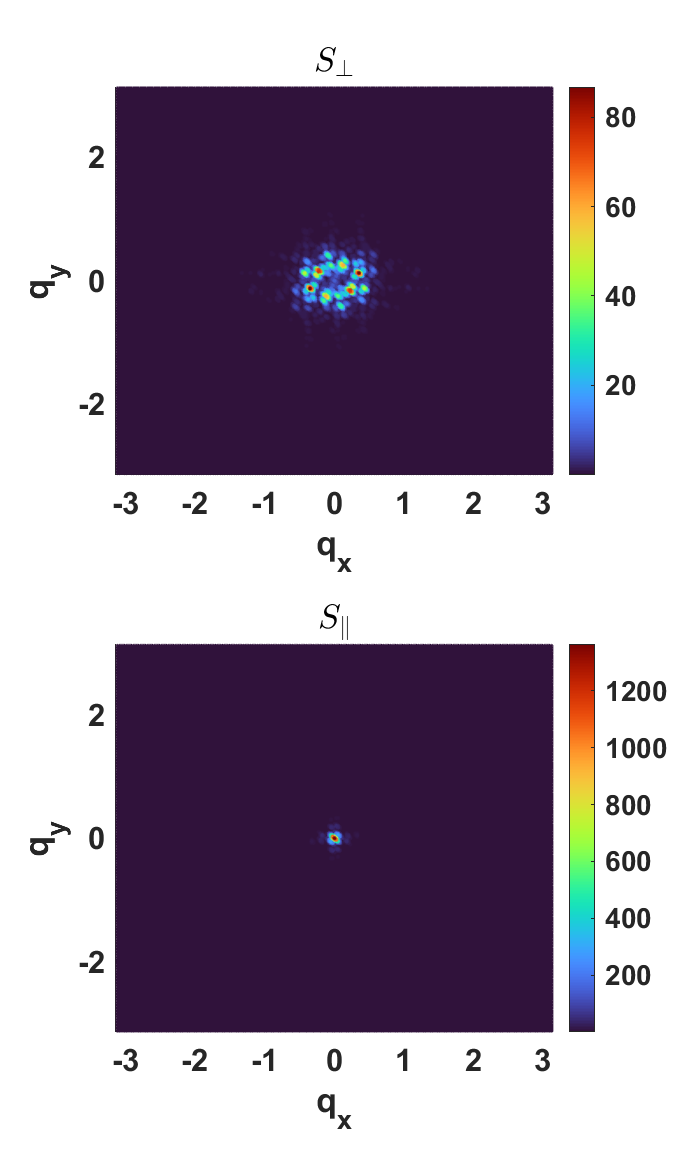}
\caption{The real-space spin configurations for $K/J=-1.2$ and $D/J=1$ for selected (a) $h/J=1.51$, (b) $h/J=1.81$ from left to the right. The middle panel displays the real-space local chirality map corresponding to that configuration. The perpendicular $S_\bot(\vec{q})$ and parallel $S_{||}(\vec{q})$ components of the static structure factor with the corresponding structure are shown in the bottom panel.}
\label{fig8}
\end{figure}

The real-space spin configurations (on the top panel) for easy-plane anisotropy $K/J=-1.2$ and $D/J=1$ can be seen in
Figs. \ref{fig7} for the selected magnetic field values (a) $h/J=0.11$, (b)$h/J=0.51$, and \ref{fig8}(a) $h/J=1.51$, (b) $h/J=1.81$. Local chirality maps ans spin structures factor maps added at columns.

In Fig. \ref{fig7}(a) $h/J=0.11$, the
spin texture is consists of a sequence of large-radius core-down (blue) merons and core-up (red) antimerons with high easy-plane anisotropy, which supports the easy-plane  (yellow) spin orientations. Here, the color transition between core-up (red) and core-down (blue) represents the $z$-component magnetization transition from $+M_z$ to $-M_z$. If these yellow arrows move clockwise towards each other, a square shape is formed surrounding the core-down (blue) merons, while for core-up (red) antimerons, they move counterclockwise, creating a square pattern.
The local chirality maps for $h = 0.11$ support the existence of meron-antimeron pairs. Note that this corresponds to the light blue region of Fig. \ref{fig2} at $(K/J,h/J)=(-1.2,0.11)$.
When we look at the structure factor in the bottom rows of Fig. \ref{fig7} (a), the perpendicular contribution is $2Q$ peak,
while the parallel $S_{||}(\vec{q})$  component has also a small peak in the center.

Fig. \ref{fig7} (b) is for the magnetic field value of $h=0.51$, and this corresponds to the yellow region of the scalar chirality map presented in Fig. \ref{fig2}.  As seen in the local chirality plot, the contribution of the red regions rises, consistent with the rising total chirality.

Spin configuration and the local chirality map are shown in Fig. \ref{fig8}(a) for the $3Q$ state at $h = 1.51$.  Only core-down merons remain in the lattice.  This texture manifests as blue dots on a red background in the local chirality map. The upward-directed spins between these merons, induced by the high magnetic field, separate the large-radius merons. Examining the perpendicular and parallel contributions of the structure factor, a weak multiple-$Q$ peak is observed near the center.
The $z$-component of the structure factor has a $1Q$ state with only a prominent peak at the center. As the diameters of the down-core merons increase and they approach each other, they cancel out the up-core antimerons between them.

The real-space spin configurations for high easy-plane anisotropy $K/J=-1.2$ and $D/J=1$ for higher magnetic field can be seen in Fig. \ref{fig8}(b) for $h/J=1.81$.  This corresponds to  the boundary between the light-blue and dark-blue regions in Fig. \ref{fig2}.
The spin texture exhibits a vortex-like structure in some regions.  Here, the $xy$ components of the spins form vortices, while the $z$ component is always positive.

The $xy$ $S_\bot(\vec{q})$ and parallel to $z$ component $S_{||}(\vec{q})$ of the structure factor can have varying peak values. The two observed peaks arise from the dynamics of domain-wall merons. In a high magnetic field, there is no creation or annihilation of merons; instead, the bimeron cluster rotates clockwise. In contrast, in a low-field situation, merons and antimerons are continuously created and annihilated, consistent with the rotational motion of the domain boundaries. The MAX phase undergoes rotations, as expected for conventional skyrmions \cite{ref31}.

In the literature, several studies have investigated the magnetic phase behavior of chiral magnets with easy-plane anisotropy and reported the emergence of various topological spin textures \cite{ref32,ref33,ref35}. In particular, theoretical studies have shown that strong easy-plane anisotropy can stabilize bimeron-type structures and related meron–antimeron configurations. These phases may emerge through transformations of conventional skyrmion lattices as the anisotropy strength increases, accompanied by the nucleation and recombination of merons and antimerons \cite{ref35}.

Isolated skyrmions and their transformation into bimeron-type textures have also been demonstrated in confined magnetic structures such as multilayer ferromagnetic disks \cite{ref36}. Furthermore, previous micromagnetic studies reported that easy-plane magnetocrystalline anisotropy can stabilize non-axisymmetric skyrmionic textures in chiral magnetic thin films \cite{ref44}.

Our results are consistent with these findings. In our simulations, increasing easy-plane anisotropy drives the system from spiral textures toward ordered  MAX phase, confirming that easy-plane anisotropy plays a key role in stabilizing bimeron-type topological textures in chiral magnetic systems.

\newpage
\section{Conclusions}\label{conclusion}

In this work, we investigated the formation of skyrmion and bimeron phases in square-lattice ferromagnetic bilayer nanostructures using Monte Carlo simulations. We identify bimerons as the primary topological agents mediating structural phase transitions in easy-plane bilayer magnets.  These simulations account for the competition between interlayer exchange interaction, the DMI, an external magnetic field, and the easy-axis/easy-plane magnetic anisotropies. The total scalar chirality map is constructed in the $(K/J-h/J)$ parameter space for the value of $D/J=1$, based on both real-space spin configurations and the evaluation of total scalar chirality. From this map and detailed investigation at some selected representative points in $(K/J-h/J)$ plane, distinct magnetic configurations including ferromagnetic states, skyrmion lattices, and MAX phase. Our results demonstrate that the transition from easy-axis to easy-plane anisotropy drives a continuous transformation from skyrmions to bimeron-type textures.

An important finding of this study is that the bilayer geometry is not equivalent to a thicker monolayer. The interlayer exchange interaction correlates the topological cores in the two layers and increases the energetic cost of local collapse of topological defects. As a consequence, bimeron and MAX phases occupy a significantly wider region of the phase diagram compared to typical monolayer behavior. The single-layer model approximates the weak-coupling limit of the bilayer system.

We further showed that the stability of these topological textures is governed by the combined effect of anisotropy and magnetic field: strong perpendicular anisotropy favors skyrmions, whereas reduced anisotropy stabilizes in-plane textures and promotes bimeron formation. This transformation occurs through a narrow, bridge-like nanotrack junction connecting the easy-axis region to the easy-plane region, as shown in previous studies in \cite{ref10,ref19,ref54}.

These results indicate that coupled magnetic layers provide an additional stabilization mechanism for bimeron-type textures and may serve as a useful platform for realizing robust topological spin structures in magnetic heterostructures. These findings deepen understanding of anisotropy-driven topological phase transitions and provide design principles for controllable skyrmion-based nanodevices.

\section*{Acknowledgment}

One of the authors (G\"ul\c{s}en Do\u{g}an) thanks to TÜBİTAK’s Directorate of Scientist
Support Programs (BİDEB) for receiving support from the 2211-C Priority Area
Domestic Doctoral Scholarship Program and TÜBİTAK for supervising her
thesis  within the subfield of ”Micro/Nano/Opto-Electromechanical Systems” and  ”Information and Communication Technologies”.

\end{document}